\documentclass[11pt]{article}
\usepackage{hyperref}
\usepackage{amsmath}
\usepackage{amsthm}
\usepackage{amsfonts}
\usepackage{algorithm}
\usepackage{algpseudocode}
\usepackage{graphicx}
\usepackage{multirow}
\usepackage{url}
\usepackage{subfig}

\title{Contrastive Matrix Completion with Denoising and Augmented Graph Views for Robust Recommendation}
\author{
		Narges Nemati\thanks{nrgs.nmt@aut.ac.ir} \textsuperscript{1}
		and
		Mostafa Haghir Chehreghani\thanks{mostafa.chehreghani@aut.ac.ir (corresponding author)} \textsuperscript{1}\\
		\textsuperscript{1}Department of Computer Engineering\\ Amirkabir University of Technology (Tehran Polytechnic)\\ Tehran, Iran}

\date{}
\begin{document}
	\maketitle
\begin{abstract}
Matrix completion is a widely adopted framework in recommender systems, as predicting the missing entries in the user--item rating matrix enables a comprehensive understanding of user preferences. However, current graph neural network (GNN)-based approaches are highly sensitive to noisy or irrelevant edges—due to their inherent message-passing mechanisms—and are prone to overfitting, which limits their generalizability.
To overcome these challenges, we propose a novel method called \textit{Matrix Completion using Contrastive Learning} (MCCL). Our approach begins by extracting local neighborhood subgraphs for each interaction and subsequently generates two distinct graph representations. The first representation emphasizes denoising by integrating GNN layers with an attention mechanism, while the second is obtained via a graph variational autoencoder that aligns the feature distribution with a standard prior. A mutual learning loss function is employed during training to gradually harmonize these representations, enabling the model to capture common patterns and significantly enhance its generalizability.
Extensive experiments on several real-world datasets demonstrate that our approach not only improves the numerical accuracy of the predicted scores---achieving up to a $0.8\%$ improvement in RMSE---but also produces superior rankings with improvements of up to $36\%$ in ranking metrics.
\\\\
\textbf{keyword:}
Matrix completion,
recommender systems,
graph neural networks (GNNs),
denoising and augmentation,
contrastive learning.
\end{abstract}

%	\end{frontmatter}
	
\section{Introduction}

With the rapid advancement of e-commerce and social media platforms in recent years, recommender systems have garnered significant attention \cite{katarya2016recent,DBLP:conf/webi/GaoZ22}. These systems provide a method for identifying user needs and predicting preferences by extracting user histories and interactions with items. Recommender system methods have been applied across various domains, including movies \cite{DBLP:journals/mlc/HwangAP23}, music \cite{DBLP:journals/evs/BakariyaSSRRM24}, and news \cite{zhu2014bayesian}. Depending on the application, recommender systems can take different forms, such as playlist generation for video and music services \cite{DBLP:journals/ijmir/SchedlZCDE18}, friend suggestions in social networks \cite{DBLP:journals/access/SuhaimB21}, and product recommendations on e-commerce sites \cite{DBLP:journals/access/AlamdariNHSD20}.

Matrix completion is a widely used setting in recommender systems \cite{DBLP:journals/cacm/CandesR12}. In this task, the rows and columns of the ratings matrix represent users and items, respectively, while the matrix entries denote user ratings for those items. By filling in the missing entries of the user-item matrix, users' preferences for {\em all} items can be predicted. Based on these predicted ratings, items are ranked for each user, and recommendations are provided accordingly.
In this setting, not only are predictions made, but mutual interactions between all users and items are also identified. As a result, the method enables a more precise and comprehensive analysis of user interests across all items.
Moreover, matrix completion techniques are essential for addressing the challenges of sparsity and accurately estimating missing values.

Various methods have been proposed to complete the user-item rating matrix. Historically, mathematical techniques \cite{DBLP:journals/computer/KorenBV09,DBLP:journals/siamjo/CaiCS10} have been predominantly used to address the matrix completion problem. However, with the rise of neural network architectures, modern approaches \cite{DBLP:conf/cikm/DuY0T23,DBLP:journals/access/ZhuFC23} now utilize deep learning to capture complex and nonlinear user-item interactions, resulting in improved prediction accuracy.
More recently, {\em contrastive learning} \cite{DBLP:conf/icml/ChenK0H20} has garnered significant attention in recommender systems. This method employs representation enhancement strategies to generate self-supervised training signals, effectively mitigating the issue of data sparsity in user-item graphs. Specifically, it utilizes graph augmentation techniques \cite{DBLP:conf/sigir/WuWF0CLX21} to create new contrastive views that closely resemble the original graph. By minimizing the loss between the original graph and the augmented views, the model learns more robust and effective node embeddings.

In general, graph neural networks (GNNs) are highly vulnerable to noisy or irrelevant edges due to their inherent message-passing mechanism. Such noise can quickly propagate through the layers, diminishing the quality of the node embeddings and ultimately degrading the recommendation performance. This issue extends even to GNN-based recommendation systems that incorporate contrastive learning techniques.
Another fundamental challenge in neural networks is overfitting, where a model learns the training data very well but fails to generalize effectively to unseen test data. Methods such as SGL~\cite{DBLP:conf/sigir/WuWF0CLX21} and SimGCL~\cite{DBLP:conf/sigir/YuY00CN22} employ random edge dropout to improve generalization; however, these techniques do not explicitly remove noise.

In this paper, we address these shortcomings by proposing a method called MCCL\footnote{MCCL stands for Matrix Completion using Contrastive Learning.}. Initially, as in IGMC \cite{DBLP:conf/iclr/ZhangC20}, we extract local neighborhood subgraphs for each interaction by extracting a 1-hop neighborhood subgraph for each node involved. Then, we employ a graph neural network in conjunction with a contrastive learning strategy. Specifically, we generate two distinct graph representations for the extracted subgraphs: one that focuses on denoising and another that generates features using a graph variational autoencoder.
For the denoising representation, we apply RGCN layers \cite{DBLP:conf/esws/SchlichtkrullKB18}. After each RGCN layer, we compute embeddings for the nodes at both ends of each edge. Then, using a multilayer perceptron (MLP), we assign an attention weight to each edge and normalize these weights to lie between 0 and 1. This process enables a meaningful differentiation among edges based on their importance. These attention weights are then incorporated during the message-passing phase in the latent space to attenuate less important connections and reinforce critical ones, thereby reducing the influence of noise present in the subgraphs.

In contrast, the graph variational autoencoder aligns the final data distribution with a standard (prior) distribution, which enables the network to generalize better to unseen data and achieve superior performance.
Finally, by employing a contrastive learning loss function during the final training phase, the two representations gradually converge, allowing the model to capture common, general patterns and significantly enhance its overall generalizability. Simultaneously, an independent loss function is defined for each representation (or sub-model) to preserve and leverage their individual strengths.

In order to evaluate our proposed model, we conduct  extensive experiments on several real-world datasets. We assess both the quality of the final rankings and the numerical accuracy of the predicted values. The results indicate that our approach not only improves the numerical accuracy of the predicted scores but also produces superior rankings. Specifically, we achieved up to a 0.8\% improvement in RMSE and up to a 36\% improvement in ranking metrics.

The rest of this paper is organized as follows. In Section 2, we present the basic concepts and definitions used throughout the paper. In Section 3, we review existing methods in the field of matrix completion for recommender systems. In Section 4, we introduce our proposed method. In Section 5, we compare our proposed method with several baseline approaches and present the results. Finally, in Section 6, we provide our conclusions and suggestions for future work.

\section{Preliminaries}

In this section, we introduce the basic concepts that underpin our study.

\paragraph{User-item interaction matrix/graph}  
The user-item interaction matrix is a central representation in recommender systems that captures the relationships between users and items. Each row corresponds to a user, each column to an item, and each matrix entry reflects the interaction—either \emph{explicit} (e.g., user ratings) or \emph{implicit} (e.g., clicks, purchases, or views). Typically, this matrix is sparse as most users interact with only a small subset of items.
Often, the interaction matrix is modeled as a \emph{bipartite graph} where users and items are represented as nodes, and any interaction between them is denoted as an edge. In this graph, nodes are divided into two disjoint sets (users and items) and edges exist solely between these sets.

\paragraph{Matrix completion}  
Matrix completion refers to the task of inferring the missing values in a partially observed matrix by identifying and exploiting underlying patterns in the available data. Once the missing entries are estimated, the completed matrix can be used to generate personalized recommendations.

\paragraph{Graph neural networks (GNNs)}  
Graph neural networks (GNNs) \cite{DBLP:conf/iclr/KipfW17,DBLP:journals/natmi/Chehreghani22,DBLP:journals/tjs/ZohrabiSC24,10.1145/3700790} are specialized neural architectures designed to process structured graph data. They function by aggregating and transforming information from a node’s neighborhood to produce low-dimensional vector representations (embeddings) for each node. Due to their multi-layer design and advanced learning strategies, GNNs are capable of capturing complex and meaningful features from graph-structured inputs.
A common variant is the \emph{Graph Convolutional Network} (GCN) \cite{DBLP:conf/iclr/KipfW17}. In a GCN, each node's embedding is iteratively updated by combining its own features with the aggregated features of its neighbors. Mathematically, this propagation rule is expressed as:
\begin{equation}
	H^{(l+1)} = \sigma\left( \tilde{D}^{-1/2} \tilde{A} \tilde{D}^{-1/2} H^{(l)} W^{(l)} \right),
\end{equation}
where \( \tilde{A} = A + I \) is the adjacency matrix with added self-loops, \( \tilde{D} \) is the corresponding degree matrix, \( H^{(l)} \) is the feature matrix at layer \( l \), \( W^{(l)} \) is the trainable weight matrix for layer \( l \), and \( \sigma \) is a nonlinear activation function (e.g., ReLU) that enables the model to learn complex representations.

\paragraph{Contrastive learning}  
Contrastive learning \cite{DBLP:conf/icml/ChenK0H20} is a self-supervised approach that trains models to distinguish between similar (positive) and dissimilar (negative) pairs of data points. By minimizing the distance between similar samples while maximizing the distance from dissimilar ones—typically through contrastive loss functions—this approach learns robust and meaningful representations without requiring labeled data.

\section{Related work}

In this section, we review existing approaches for matrix completion in recommender systems. These methods can be broadly classified into two categories: \textbf{mathematical methods} and \textbf{neural network-based methods}.
Below, we discuss these approaches in detail.

\subsection{Mathematical methods}

\subsubsection{Non-negative matrix factorization}
Lee et al.~\cite{DBLP:conf/nips/LeeS00} introduced non-negative matrix factorization (NMF) as a dimensionality reduction technique. NMF approximates a large and sparse user-item interaction matrix \(R\) by factorizing it into two low-rank non-negative matrices:
\begin{equation}
	R \approx W H^T,
\end{equation}
where \(W\) denotes the latent user factor matrix and \(H\) represents the latent item factor matrix. The inherent non-negativity constraints ensure that the latent factors are interpretable.

Following the initial formulation of matrix factorization, several extensions have been proposed to improve its performance~\cite{DBLP:journals/mpc/WenYZ12,DBLP:conf/csai/Wang21}. For instance, Deng et al.~\cite{DBLP:journals/tsmc/DengRWZG23} introduced \emph{probabilistic matrix factorization} (PMF), a Bayesian approach that models user-item interactions via probabilistic latent representations. In PMF, user and item latent factors are treated as random variables with Gaussian priors, and the observed ratings are assumed to be sampled from a Gaussian distribution centered on the dot product of the latent vectors \(\mathbf{p}_u\) and \(\mathbf{q}_i\), with variance \(\sigma^2\). This framework maximizes the likelihood of the observed data while regularizing the latent factors with their respective priors.

\subsubsection{Methods based on singular value decomposition}
Cai et al.~\cite{DBLP:journals/siamjo/CaiCS10} introduced the Singular Value Decomposition (SVD) method, which decomposes the user-item rating matrix into three components: a user feature matrix, a diagonal singular value matrix, and an item feature matrix. Specifically, for a matrix \(A\), SVD yields:
\[
A = U \Sigma V^T,
\]
where \(U\) and \(V\) capture the latent factors for users and items, respectively, and \(\Sigma\) contains the singular values that inform the importance of these factors.

Several extensions have subsequently been proposed~\cite{DBLP:conf/iccse2/JiHSWM16,DBLP:conf/kdd/Koren08}. Koren et al.~\cite{DBLP:conf/kdd/Koren08} enhanced standard SVD by incorporating implicit feedback (such as clicks, views, and purchases) into the user representations, thereby capturing more nuanced user preferences. Similarly, Zhou et al.~\cite{DBLP:conf/aaim/ZhouWSP08} introduced the \emph{alternating least squares} (ALS) method, which approximates the user-item rating matrix by factorizing it into two lower-dimensional matrices, with each user and item represented by a vector in a latent space. The predicted rating is then computed as the dot product of these vectors.

\subsubsection{Spectral regularization and nuclear norm minimization}
Spectral regularization~\cite{DBLP:journals/jmlr/MazumderHT10} imposes a penalty directly on the singular values of a matrix rather than on the latent factors. By shrinking smaller singular values toward zero during optimization, the method effectively favors solutions with a lower rank. Since directly minimizing the rank is NP-hard, \emph{nuclear norm minimization}~\cite{DBLP:journals/cacm/CandesR12} offers a convex alternative by minimizing the sum of the singular values (the nuclear norm) while enforcing consistency with the observed entries.

\subsection{Neural network-based methods}

\subsubsection{Utilizing Autoencoders}
Autoencoders~\cite{hinton2006reducing,DBLP:journals/corr/KingmaW13} are neural networks that learn to represent data in a compressed, lower-dimensional space and then reconstruct the original input. In recommender systems, where the user-item matrices are typically low-rank, the hidden layers of an autoencoder capture latent factors. The AutoRec model~\cite{DBLP:conf/www/SedhainMSX15} leverages this framework to refine representations of user-item rating matrices and can be implemented in both user-based and item-based fashions. Additionally, Shanian et al.~\cite{DBLP:conf/www/HeLZNHC17} proposed a neural network-based matrix factorization method that replaces the standard dot product with a deep neural network to better capture complex user-item interactions.

\subsubsection{Recommender systems using contrastive learning}
Recent approaches have increasingly employed contrastive learning to enhance user-item interaction predictions~\cite{DBLP:conf/sigir/YuY00CN22,DBLP:conf/kdd/Jiang0H23,DBLP:journals/eswa/WeiXZMP24}. These methods typically construct two distinct views of the user-item graph, using a contrastive loss to align their representations in conjunction with additional loss functions to improve prediction accuracy. For example, Shin et al.~\cite{DBLP:journals/ijon/MoZHQL25} proposed a framework where one view is generated by randomly removing edges and processing the graph with convolutional layers, while the other employs an attention-enhanced GCN on the bipartite graph. The two views are then encoded using a variational graph autoencoder (VGAE)~\cite{DBLP:journals/corr/KipfW16a} to produce the final node embeddings. Similarly, Shaheng et al.~\cite{DBLP:conf/iclr/Cai0XR23} constructed two graph views—one via GCN and the other through SVD—and applied a contrastive loss alongside a prediction loss for alignment.

\subsubsection{Leveraging large language models}
Recent recommendation methods have also explored the use of large language models (LLMs) to derive meaningful representations from auxiliary data such as user reviews, product descriptions, and other metadata. Benefiting from advanced natural language processing capabilities, LLMs can extract semantic and fine-grained features from text, which enhances personalization. For instance, Xiang et al.~\cite{DBLP:conf/mlmi2/XiangHGYZ24} proposed a multimodal matrix factorization approach that integrates traditional matrix factorization with LLMs (using architectures akin to RoBERTa) and computer vision techniques to extract rich latent features from textual and visual data.

\subsection{Matrix completion using graph neural networks}

Berg et al.~\cite{DBLP:journals/corr/BergKW17} introduced matrix completion with graph convolutional networks (GCNs) by modeling the rating matrix as a bipartite graph, where users and items form distinct sets of nodes connected via interaction edges. A GCN is subsequently employed to generate latent embeddings through message passing across this graph.
Inductive Matrix Completion based on Graph Neural Networks (IGMC)~\cite{DBLP:conf/iclr/ZhangC20} extends this concept by leveraging GNNs to capture higher-order relationships between users and items while preserving inductive capabilities—thus enabling the model to generalize to unseen user-item pairs during inference. In IGMC, the user-item interaction matrix is represented as a bipartite graph with weighted edges derived from observed ratings. The method focuses on extracting local subgraphs surrounding each observed interaction, which are then processed by a GNN to predict the corresponding ratings.
Inductive Matrix Completion using a Graph Autoencoder~\cite{DBLP:conf/cikm/0005ZTZHD021} adopts a similar strategy: it extracts local subgraphs around user-item interactions and applies a graph autoencoder to process the subgraph structures. A graph convolutional network (GCN) encodes the node embeddings by learning local graph patterns, thereby capturing the intrinsic structure of the data.

Ziaee et al.~\cite{DBLP:journals/kais/ZiaeeRN24} proposed a movie recommendation method that exploits heterogeneous graph structures. In their approach, textual summaries of movies from Wikipedia are processed using the BERT model to extract 768-dimensional embeddings for movie titles. Using cosine similarity, movie pairs with a similarity score above 95\% are connected with a weight of one. Additionally, genre embeddings of size 32, extracted via BERT and refined with Jaccard similarity, are incorporated. Similar to IGMC, subgraphs are then extracted and initial node embeddings are assigned. The model employs a graph autoencoder with two GNN layers for encoding and aggregates user and item embeddings during decoding, with a multi-layer perceptron (MLP) producing the final predictions.
Han et al.~\cite{DBLP:journals/eaai/HanKHY24} proposed a method to enhance item recommendations by extracting keywords from user reviews. In this approach, keywords derived from all reviews provided by a user and those received by an item are used as node features. Analyzing keyword overlaps between nodes introduces additional edges into the initial user-item bipartite graph. Following a strategy similar to IGMC, subgraphs are extracted from the heterogeneous graph and processed by a relational graph convolutional network (R-GCN) to generate improved node embeddings.

HetroFair \cite{DBLP:journals/corr/abs-2402-03365} is a GNN-based recommender that improves fairness for item providers. It uses fairness-aware attention and heterophily feature weighting to reduce popularity bias.
The authors of \cite{DBLP:journals/corr/abs-2502-15699} developed a GNN-based recommendation model that disentangles popularity and quality,
by introducing an edge classification technique.
Their method alsouses cost-sensitive learning
to adjust the misclassification penalties.
The authors of \cite{akhlaghi2025adaptivelongtermembeddingdenoising}
proposed a recommendation system in the dynamic setting,
that combines an item-item graph with GCNs to capture short-term interactions and uses GRUs with attention for long-term user preferences. It incorporates denoising and data augmentation techniques along with an adaptive weighting strategy to balance these signal.

\section{Methodology}

In this section, we first formally define our studied problem.
Then, we present different components of our proposed MCCL method\footnote{Our implementation of MCCL is publicly available at \url{https://github.com/NargesNemati/MCCL}.}.
%The overall architecture of the ALDA4Rec model id depicted in Figure \ref{fig:frame}.

\subsection{Problem definition}

The problem of matrix completion in recommender systems is defined as follows.
A set of users $U = \{u_1, u_2, \dots, u_n\}$ and a set of items $I = \{i_1, i_2, \dots, i_m\}$ are given, resulting in $n$ users and $m$ items. Users are placed in the rows and items in the columns of a matrix.
User-item interactions are represented by a partially observed rating matrix $R \in \mathbb{R}^{m \times n}$, where $R_{ui}$ indicates the rating given by user $u$ to item $i$. This user-item interaction matrix $R$ is often very sparse, meaning that many of its entries are empty. The goal of the matrix completion problem is to fill these empty entries.

In many cases, this user-item interaction matrix is transformed into a bipartite graph. A bipartite graph $G = (V, E, R)$ is defined, where $V = U \cup I$ is the set of nodes, with $U$ representing the users and $I$ the items. The edges $E \subseteq U \times I$ connect users and items that have interactions. These edges can be weighted, with weights $R_{ui}$ indicating the observed ratings of users for items. The goal, in this scenario, is to predict the weights of edges between users and items.

\subsection{Proposed method}

\begin{figure}[h]
	\centering
	\includegraphics[width=0.7\textwidth]{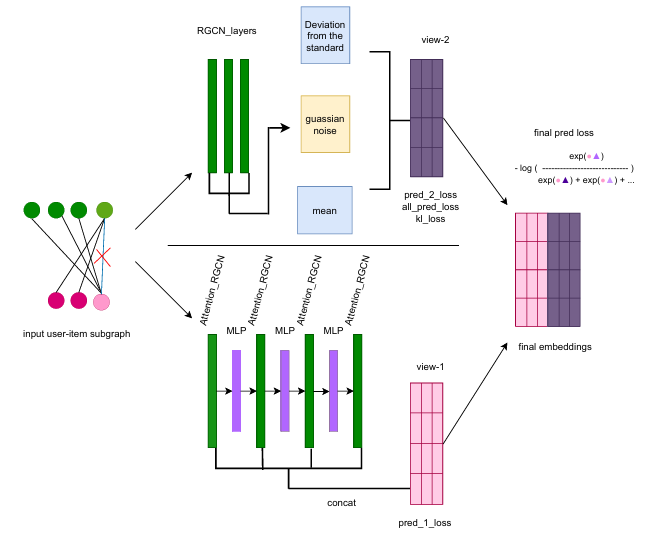}
	\caption{Overall framework of the proposed MCCL model.}
	\label{model}
\end{figure}

Figure \ref{model} illustrates an overview of the proposed method. The input to the model is a subgraph extracted from the user-item bipartite graph. This subgraph is constructed around the target user and item nodes, with the interaction edge of the target being removed. 
The upper section of the model comprises a variable graph autoencoder, while the lower section incorporates a noise removal model utilizing attention mechanisms. Finally, the embeddings from these two models are combined to make the final predictions. 
Details of the proposed method will be elaborated in the following sections.

\subsubsection{Extracting subgraphs}

Similar to the IGMC model \cite{DBLP:conf/iclr/ZhangC20}, we extract subgraphs from the original bipartite graph. In this case, we consider a user $u$ and an item $i$. All nodes in the neighborhood of one of these two target nodes are included.
The edge weights corresponding to interactions in the initial interaction matrix are set to the rating given by the user for the item. These ratings typically range from one to five. For the nodes in the subgraphs, initial embeddings are assigned, and these embeddings are determined based on the original user-item bipartite graph. We remove the initial interaction edge from the subgraph, and the ultimate objective of our method is to predict the weight of this edge.

For the initial embedding of nodes, the value of 1 is assigned to the target user node, and the value of 2 is assigned to the target item node. Subsequently, for other neighbors, the value assigned to user nodes is $1 + i \times 2$, where $i$ represents the distance from the target item node. Since we consider a distance of 1, this value becomes 3. For item nodes, the value assigned is $i \times 2$, where $i$ represents the distance from the target user node. Similarly, for a distance of 1, this value becomes 4.
These values are then converted into one-hot encoded vectors for decoding. The initial embeddings for the nodes are as follows:
\begin{itemize}
	\item Target user node: $[1, 0, 0, 0]$
	\item Target item node: $[0, 1, 0, 0]$
	\item Neighboring user nodes: $[0, 0, 1, 0]$
	\item Neighboring item nodes: $[0, 0, 0, 1]$
\end{itemize}

\subsubsection{Contrastive learning}

We use the contrastive learning approach, where  two distinct views are created for each subgraph. One view is generated using a variational graph autoencoder, while the other is constructed by denoising the initial subgraph.
For each subgraph, we thus have two distinct views, from which node embeddings are derived. The network is then trained using the contrastive loss function along with other loss functions, as explained in Section \ref{sec:loss}. This process enables the models to observe and leverage features from each other, enhancing their representation.
Finally, the embeddings from both views are combined, resulting in an overall improvement in graph representation.

\subsubsection{Constructing a denoised graph using attention mechanisms}

In the initial subgraphs, there may exist noisy data that can degrade the performance of the model. 
For the first view, since each edge can have different types based on its weight (e.g., varying ratings given by users to items), we utilize a relational graph convolutional network (RGCN) \cite{DBLP:conf/esws/SchlichtkrullKB18}. 
To perform noise removal, an attention value is assigned to each edge, which is multiplied by the weight that the RGCN assigns to that edge during message passing.

In the first layer, we use a standard relational graph convolutional network (RGCN). After each layer, the embeddings obtained for the nodes in the previous layer are passed through the relational graph convolutional network. For each edge in the subgraph, the embeddings of the user and item at the two endpoints of the edge are concatenated and fed into a multi-layer perceptron (MLP), which outputs a single value for each edge.

To ensure noise removal and to create a significant distinction between the remaining and removed edges, these obtained values for the edges are normalized to the range of [0, 1]. This normalization determines the attention weight for each edge. During message passing in the next layer, this attention weight is multiplied by the weight determined by the graph convolutional network for each edge in the subgraph.
As a result, the attention-weighted values in the weight matrix for the edges are constrained to the range of $[0, 1]$, which reduces the influence of noisy values and ensures that the attention mechanism effectively removes noise.

Finally, we use the dot product between the target nodes, and the result of this operation for the user and item target nodes represents the model's predicted score for the target edge. The relationships are formulated as follows:
\[
\text{att}^{(l)}_{u,i} = \text{MLP} \left( h^{(l)}_u \Vert h^{(l)}_i \right)
\]
where \( h^{(l)}_u \) and \( h^{(l)}_i \) denote the embeddings of user \( u \) and item \( i \), respectively, at the l-th layer of the relational graph convolutional network.
These two embeddings are concatenated and fed into a multi-layer perceptron (MLP). Subsequently, for normalization of these values, we use the following formula:
\[
\alpha^{(l)}_{u,i} = \frac{\text{att}^{(l)}_{u,i} - \min(\text{att}^{(l)}_{U',I'})}{\max(\text{att}^{(l)}_{U',I'}) - \min(\text{att}^{(l)}_{U',I'})}
\]
where $U'$ includes all users and $I'$ includes all items.
In this way, the attention weight of each edge is obtained.
Then, we use the following equation for message passing:
\begin{equation}
h^{(l+1)}_v = \sigma
\left(
\sum_{r}
\left(
\sum_{j \in N^{r}_v}
\alpha^{(l)}_{v,j} W^{(l)}_r h^{(l)}_j
\right)
+ \alpha^{(l)}_{v,v} W^{(l)}_0 h^{(l)}_v
\right)
\end{equation}
In this formula, \( j \) represents all neighboring nodes of \( v \) by edge weight \(r\) between \(v\) and \(j\). $W^{(l)}_r$ is the weight matrix assigned to edges of weight r in the l-th layer of the RGCN, and $W^{(l)}_0$ is the weight matrix for the self-loops in that same layer. These matrices are learned during training. Finally, each attention score is multiplied by its corresponding RGCN edge weight to account for differences among edges and reduce noise.

\subsubsection{Graph construction using a variational graph autoencoder} 

To create the second view of the graph, we utilize the variational graph autoencoder \cite{DBLP:journals/corr/KipfW16a}.
%In Chapter 2, the variational graph autoencoder was defined, and here we provide a more detailed explanation of the model and the specifics of our proposed approach. 
Variational graph autoencoders are designed for unsupervised learning on graphs and are particularly useful for tasks such as link prediction. Instead of mapping an input to a deterministic hidden vector, variational graph autoencoders model each node with a probabilistic latent variable, enabling variability in the learned representations.
The encoder processes the node features \( X \) along with the graph adjacency matrix \( A \). For each node, it outputs the parameters of the latent distribution (the mean \( \mu_i \) and standard deviation \( \sigma_i \)) as:
\[
z_i \sim \mathcal{N}(\mu_i, \sigma_i^2).
\]

For the encoder, we first define two layers of a RGCN \cite{DBLP:conf/esws/SchlichtkrullKB18}. Then, using a multi-layer perceptron (MLP), we extract the mean (\( \mu \)) and the logarithm of the standard deviation (\( \log(\sigma) \)). Subsequently, Gaussian noise is generated based on the length of deviation (\( \log(\sigma) \)) vector for each node.
Finally, the final embeddings of the encoder are obtained as follows:
\[
Z = \mu + \epsilon \odot \sigma 
\]
Where \( \sigma \) represents the standard deviation, obtained using the formula \( \exp(\log(\sigma)) \). The value \( \epsilon \) corresponds to Gaussian noise, and finally, \( z \) represents the final node embeddings produced by the encoder.

The decoder reconstructs the graph structure from the latent variables. This is achieved by calculating the probability of an edge between two nodes as the dot product of their latent embeddings. 
In this step, we reconstruct all edges in the subgraphs and also predict the target edge that were removed from the subgraph.

\subsubsection{Prediction head}

For the final model, the embeddings obtained from the two previous models are collected for the target users and items. Finally, for each target edge, the final embeddings of the user and the item at the two endpoints of the edge are concatenated and fed into a multi-layer perceptron (MLP). The output of this network is a single value, which represents the final prediction of our model.

\subsubsection{Loss functions}
\label{sec:loss}

Distinct loss functions are defined for each of the two graph views. Additionally, a loss function is considered for contrastive learning, and another is defined for final predictions.
For the first view, which involves noise removal using the attention mechanism, we use the MSE (Mean Squared Error) loss function, defined as:
\begin{equation}
	{L_{view-1}} = \frac{1}{N} \sum_{i=1}^{N} \left( r_i - \hat{r}_i \right)^2
\end{equation}
where \( N \) is the number of observed interactions, \( r_i \) is the actual rating for the \( i \)-th interaction, and \( \hat{r}_i \) is the predicted rating for the \( i \)-th interaction.

For the second view, which uses the variational graph autoencoder, three loss functions are utilized. The loss function for target edge prediction is formulated as:
\begin{equation}
	{L_{pred}} = \underbrace{ \frac{1}{N} \sum_{i=1}^{N} \left( r_i - \hat{r}_i \right)^2} _ {\text{target edges loss}}
\end{equation}
where \( N \) represents the number of user-item actual interactions whose edges have been removed from the input subgraph, \( r_i \) is the actual rating for the \( i \)-th interaction, and \( \hat{r}_i \) is the predicted rating for the \( i \)-th interaction.

The second loss function is defined as follows:
\begin{equation}
	{L_{rec}} = \underbrace{ \frac{1}{M} \sum_{j=1}^{M} \left( w_j - \hat{w}_j \right)^2}_{\text{subgraphs reconstruction loss}}
\end{equation}
where \( M \) is the total number of edges in the input subgraph, \( w_j \) represents the actual weight, and \( \hat{w}_j \) denotes the predicted weight for the \( j \)-th edge.
The third loss function is defined as follows:
\begin{equation}
	- \underbrace{D_{KL}\left(q(z \mid\\\, A p(z)\right)}_{\text{regularization term}}
\end{equation}
The KL divergence measures how one probability distribution diverges from a reference distribution. In the context of a variational graph autoencoder, it acts as a regularization term that encourages the learned latent variable distribution to remain close to an initial distribution, typically a standard normal distribution.
This term is calculated as follows:
\begin{equation}
	D_{KL}\Bigl(q(z \mid A) \,\Bigl\|\, p(z)\Bigr) = -\frac{1}{2}\sum_{i=1}^{d} \Bigl(1 + \log(\sigma_i^2) - \mu_i^2 - \sigma_i^2\Bigr)
\end{equation}
where \( d \) represents the dimensions of the latent space, \( \mu_i \) is the mean, and \( \sigma_i^2 \) is the variance for each node. This value is computed for all nodes and then either averaged or summed over all nodes. We use averaging:
\begin{equation}
	{L_{KL}} = \frac{1}{V} \sum_{i=1}^{V} D_{KL-i}
\end{equation}
where \( V \) is the number of nodes, and \(L_KL\) is added to balance learning and prevent overfitting within the overall loss function.

The second view's loss function is defined as follows:
\begin{equation}
	{L_{view-2}} = L_{pred} +  \alpha \times L_{rec} + \beta \times L_{KL}.
\end{equation}
Hyperparameters \( \alpha \) and \( \beta \) control the influence of \( L_{\text{rec}} \) and \( L_{\text{KL}} \) on the total loss function, respectively.

For the final model, a loss function is calculated to measure the difference between the predicted and actual values, as well as a loss function for contrastive learning. The contrastive learning loss is defined as:
\begin{equation}
	L_s(user) = \sum_{u_i \in U} -\log \frac{\exp \left( s(e'_i, e''_i) / \tau \right)}{\sum_{u_{i'} \in U} \exp \left( s(e'_i, e''_{i'}) / \tau \right)}
\end{equation}
where \( e'_i \) represents the output embedding from the second view for user \( i \), \( e''_i \) represents the output embedding from the first view for user \( i \), and \( \tau \) is a hyperparameter. This formula maximizes the similarity of the embedding for user \( i \) in the two views relative to the similarity of the embedding for user \( i \) in the first view to all users in the second view.

Similarly, a comparable loss function is calculated for items, and the two losses are summed. For the final output of the model, the following prediction loss function is defined:
\begin{equation}
	{L_{pred-final}} = \frac{1}{N} \sum_{i=1}^{N} \left( r_i - \hat{r}_i \right)^2
\end{equation}
This loss function is computed similarly to the prediction loss for the previous two views. For the final model, we have:
\begin{equation}
	L_{final-model} = L_{pred-final} + \lambda \times (L_s(user) + L_s(item))
\end{equation}
where \( \lambda \) is a hyperparameter that determines the influence of the contrastive learning loss on the overall loss function.

Finally, for the combined model, all loss functions are summed, and the total loss function is defined as:
\begin{equation}
	L = L_{view-1} + L_{view-2} + L_{final-model}
\end{equation}
This way, all loss functions are aggregated together.

\subsection{Algorithm}

In this section, we present our proposed algorithm in the form of pseudocode for training a batch, followed by an analysis of its time complexity.

\paragraph{Algorithm \ref{alg:view-1}} In this algorithm, we examine the construction of the first view within a batch.
To do so, we first identify the source and destination nodes of the edges (Lines 1-2). Then, a layer of a RGCN is applied (Line 3). For the next three layers, the embeddings obtained from the previous layer for the two endpoints of each edge are concatenated and passed through a multi-layer perceptron (MLP) (Line 5). Normalization is then applied (Line 6). Subsequently, the subgraphs and the weights obtained for each edge are fed into a relational graph convolutional layer with attention (the attention weights are applied during the message-passing phase) (Line 7).
Finally, a dot product is performed for the target user and item nodes (Lines 9-11). The output of this model includes the predicted target edges and the embeddings obtained for the nodes.

\begin{algorithm}
		\caption{Constructing the first view.\label{alg:view-1}}
			\setlength{\baselineskip}{0.8\baselineskip}
			\begin{algorithmic}[1]
				\Require \( sub\_graphs = (\mathcal{V}, \mathcal{E}, W) \)
				\State \( src\_nodes \gets \mathcal{E}.\text{sources} \)
				\State \( dst\_nodes \gets \mathcal{E}.\text{destinations} \)
				\State \( H \gets \text{RGCN}(sub\_graphs) \)
				\For{\( i = 0 \) \textbf{to} \( 2 \)}
				\State \begin{minipage}[t]{\dimexpr\linewidth-\algorithmicindent}
					\( first\_att \gets \text{MLP}(\text{concat}(H[src\_nodes], \quad H[dst\_nodes], \text{dim}=1)) \)
				\end{minipage}
				\State \( attention\_weight \gets \text{normalization}(first\_att) \)
				\State \( H \gets \text{RGCN\_with\_attention}(G, attention\_weight) \)
				\EndFor
				\State \( H[u] \gets sub\_graphs.node\_data[:0] == 1 \)
				\State \( H[i] \gets sub\_graphs.node\_data[:1] == 1 \)
				\State \( prediction_1 \gets H[u] . H[i] \)
				\State \textbf{Output:}  $prediction_1$, $H$
			\end{algorithmic}
	\end{algorithm}

\paragraph{Algorithm \ref{alg:view-2}} In this algorithm, we examine the construction of the second view within a batch.
To do so, we first identify the source and destination nodes of the edges in the input subgraphs (Lines 1-2). Next, three layers of a RGCN are applied (Lines 3-5). 
Then, using a multi-layer perceptron (MLP), the mean and the logarithm of the standard deviation are computed for the node embeddings (Lines 6-7). Gaussian noise is generated based on the dimensions of the standard deviation (Lines 8). 
The updated embeddings are then obtained (Line 9). The edge weights of the subgraph are calculated using the new embeddings and the dot product operation and the predicted weights for the target edges that were removed from the subgraph are computed using the new embeddings (Lines 10-11).

	\begin{algorithm}
		\caption{Constructing the second view.\label{alg:view-2}}
			\setlength{\baselineskip}{0.8\baselineskip}
			\begin{algorithmic}[1]
				\Require \( sub\_graphs = (\mathcal{V}, \mathcal{E}, W) \)
				\State \( src\_nodes \gets \mathcal{E}.\text{sources} \)
				\State \( dst\_nodes \gets \mathcal{E}.\text{destinations} \)
				\For{\( i = 0 \) \textbf{to} \( 2 \)}
				\State \(H \gets \text{RGCN}(sub\_graphs) \)
				\EndFor
				\State \( \text{mean} \gets \text{MLP}(embeddings) \)
				\State \( \text{std} \gets \text{MLP}(embeddings) \)
				\State \( \text{gaussian\_noise} \gets \text{SampleNormal}(0, 1, \text{std.shape}[0]) \)
				\State \( H \gets \text{mean} + \text{gaussian\_noise} \times \exp(\text{std}) \)
				\State \( prediction_2 \gets (H[point\_users] . H[point\_items] \)
				\State \( all\_predictions_2 \gets H[all\_users] . H[all\_items] \)
				\State \begin{minipage}[t]{\dimexpr\linewidth-\algorithmicindent}  \textbf{Output:} \( prediction_2,\; all\_predictions_2,\; \text{mean},\; \text{std},\; H \)
				\end{minipage}
			\end{algorithmic}
	\end{algorithm}

\paragraph{Algorithm \ref{alg:final-model}} 
In this algorithm, we examine the construction of the contrastive learning model and the calculation of the contrastive learning loss within a batch.
We first define a function to calculate the contrastive learning loss for the final embeddings derived from the two previous views (Lines 1-12). Next, the final embeddings of the users from the two different views are concatenated to obtain the users' final embeddings. The same operation is performed for the items. Then, these embeddings are concatenated and passed through a multi-layer perceptron (MLP) and a sigmoid non-linear activation function to generate the final predictions (Lines 13-18).
In the end, the loss functions for both views, the contrastive learning loss function, and the prediction loss function of the final model are calculated and combined with their respective weights.

\begin{algorithm}
		\caption{The final model.\label{alg:final-model}} 
			\setlength{\baselineskip}{0.8\baselineskip}
			\begin{algorithmic}[1]
				\Require $H[view_1]$, $H[view_2]$				
				\Function{ContrastiveLoss}{$z1, z2, \tau$}
				\State $similarities \gets \frac{z1 \times (z2)^T}{\tau}$
				\State $n \gets \text{number of rows in } similarities$
				\State $loss \gets 0$
				\For{$i \gets 0$ to $n-1$}
				\State $pos\_sim \gets similarities[i][i]$
				\State $neg\_sim \gets \text{concatenate}( similarities[i][0:i],\, similarities[i][i+1:n])$
				\State $denom \gets \sum \exp(neg\_sim)$
				\State $loss \gets loss - \log\!\left(\frac{\exp(pos\_sim)}{denom + \epsilon}\right)$ \Comment{$\epsilon$ is a small constant to avoid division by zero}
				\EndFor
				\State \Return $\frac{loss}{n}$
				\EndFunction
				
				\State $user\_loss \gets \Call{ContrastiveLoss}{H[view_1\_users],\, H[view_2\_users],\, \tau}$
				\State $item\_loss \gets \Call{ContrastiveLoss}{H[view_1\_items],\, H[view_2\_items],\, \tau}$
				\State $contrastive\_loss \gets \frac{user\_loss + item\_loss}{2}$
				\State $H[u] \gets \text{Concatenate}(H[view_1\_users],\; H[view_2\_users])$
				\State $H[i] \gets \text{Concatenate}(H[view_1\_items],\; H[view_2\_items])$
				\State $preds \gets \text{sigmoid}\Bigl(\text{MLP}\Bigl(\text{Concatenate}(H[u],\, H[i])\Bigr)\Bigr)$
				\State \textbf{Output:} \( \text{preds},\; \text{contrastive\_loss} \)				
			\end{algorithmic}
	\end{algorithm}

\subsubsection{Complexity analysis}

We define
	\( L_1 \) as the number of RGCN layers,
	\( L_2 \) as the number of layers in each MLP,
	\( d_{in} \) as the size of the input node features, and
	\( d_{out} \) as the size of the output node features.
The computational complexities for the different components of MCCL are:
\begin{itemize}
	\item
	\textbf{Noise removal component:} Each RGCN layer has a time complexity of 	
$	O\left(|E| \cdot d_{in} \cdot d_{out}\right),$	
	so the overall complexity is 
$
	O\left(L_1 \cdot |E| \cdot d_{in} \cdot d_{out}\right).
$
	
	\item
	\textbf{Variational graph autoencoder:} The RGCN layers here incur the same complexity:	
$
	O\left(L_1 \cdot |E| \cdot d_{in} \cdot d_{out}\right).
$	
	\item \textbf{MLP:} Processing the nodes results in a time complexity of 
$
	O\left(L_2 \cdot d_{in} \cdot d_{out} \cdot |V|\right).
$
	\item \textbf{Final model:} Each edge is processed by an MLP, leading to a time complexity of 	
$
	O\left(L_2  \cdot d_{in} \cdot d_{out} \cdot |E| \right).
$
Note that the maximum number of layers in the MLP is 2.
\end{itemize}
Thus, the total time complexity is:
\[
O\left( L_1 \cdot |E| \cdot d_{in} \cdot d_{out} + L_2 \cdot d_{in} \cdot d_{out} \cdot \left(|V|+|E|\right) \right).
\]
Assuming that the user-item graph is connected (i.e., \(|V| = O(|E|)\)), this simplifies to:
$
O\left( (L_1 + L_2) \cdot |E| \cdot d_{in} \cdot d_{out} \right).
$
This time complexity is equivalent to that of the IGMC method,
whose subgraph extraction technique is exploited by our method.

%In the IGMC method where we extracted subgraphs in this way, we used RGCN layers for training and an MLP for edge prediction. Our method has a higher time complexity compared to this method by the following amount:
%\[
%O(L_2 \cdot d_{in} \cdot d_{out} \cdot |V|).
%\]

\section{Experiments}
In this section, we first provide an overview of the datasets, baseline models, and evaluation metrics used in our study. We then present our extensive experiments, detailing the evaluation settings and analyzing the impact of various factors on model performance.

\subsection{Evaluation metrics}

In our study, we assess the recommender system performance using five metrics:

\begin{enumerate}
	\item \textbf{Root Mean Square Error (RMSE)}:  
	This metric measures the average magnitude of prediction errors between the actual and predicted ratings. Lower RMSE values indicate better prediction accuracy.

	\[
	\text{RMSE} = \sqrt{\frac{1}{n} \sum_{i=1}^{n} (\hat{y}_i - y_i)^2},
	\]

	where \( y_i \) is the actual rating, \( \hat{y}_i \) is the predicted rating, and \( n \) represents the number of predictions.
	
	\item \textbf{Mean Reciprocal Rank considering Ratings (MRR-rating)}:  
	This metric evaluates ranking quality by computing the average inverse rank of the first highly rated item within the top-\(N\) recommendations for each user.

	\[
	\text{MRR@N-rating} = \frac{1}{|Q|} \sum_{i=1}^{|Q|} \frac{1}{\text{rank}_i},
	\]

	where \( \text{rank}_i \) is the rank of the first highly rated item for user \( i \) and \(|Q|\) is the total number of users.
	
	\item \textbf{Normalized Discounted Cumulative Gain considering Ratings (NDCG-rating)}:  
	This metric assesses ranking quality by taking into account both the relevance scores and the positions of items in the ranking. First, we define:

	\[
	\text{DCG@N} = \sum_{i=1}^{N} \frac{rel_i}{\log_2(i+1)},
	\]

	where \( rel_i \) is the relevance score of the item at position \( i \). The Ideal DCG (IDCG) is computed for the best possible ranking:

	\[
	\text{IDCG@N} = \sum_{i=1}^{N} \frac{ideal\_rel_i}{\log_2(i+1)}.
	\]

	Then, NDCG is given by:

	\[
	\text{NDCG@N-rating} = \frac{\text{DCG@N}}{\text{IDCG@N}}.
	\]

	\item \textbf{Mean Reciprocal Rank with Positive and Negative Labels (MRR-ranking)}:  
	Here, the last interaction of a user is treated as the positive item (with a fixed relevance score), while a set of 99 negative items is generated. The metric is computed as:

	\[
	\text{MRR@N-ranking} = \frac{1}{|Q|} \sum_{i=1}^{|Q|} \frac{1}{\text{rank}_i + 1},
	\]

	where \( \text{rank}_i \) is the rank of the positive item for user \( i \) among the top-\(N\) recommendations.
	
	\item \textbf{Normalized Discounted Cumulative Gain with Positive and Negative Labels (NDCG-ranking)}:  
	Similar to NDCG-rating, this metric evaluates ranking quality in a scenario with one positive and several negative items:

	\[
	\text{NDCG@N-ranking} = \frac{1}{|Q|} \sum_{i=1}^{|Q|} \frac{1}{\log_2(\text{rank}_i + 1)},
	\]

	where \( \text{rank}_i \) corresponds to the rank position of the positive item for user \( i \) among the top-\(N\) recommendations.
\end{enumerate}

\subsection{Baseline models}

This section provides a brief description of the baseline models and their proposed methodologies.

\begin{itemize}
	\item \textbf{IGMC \cite{DBLP:conf/iclr/ZhangC20}}: For each interaction in the bipartite user-item graph, a one-hop local subgraph is extracted, which contains users and items within a one-hop distance of the target interaction. The edge corresponding to the target interaction is then removed, and the resulting subgraph is fed into a relational graph convolutional network for predicting the missing edge.
	
	\item \textbf{SGL \cite{DBLP:conf/sigir/WuWF0CLX21}}: This method leverages contrastive learning by generating multiple graph views. One view is created by randomly removing a subset of nodes and their connected edges (reducing over-reliance on high-degree nodes), while another view is formed by randomly removing certain edges. Additionally, random walks are used at each layer to remove edges. The embeddings from these perturbed views are aligned using a contrastive loss function, and a separate loss (e.g., BPR loss) is employed for predictions.
	
	\item \textbf{SimGCL \cite{DBLP:conf/sigir/YuY00CN22}}: Similar to SGL, this approach applies contrastive learning by generating two distinct graph views via random edge removal. The embeddings from these views are aligned through a contrastive loss function, which is combined with a prediction loss (BPR loss) to improve overall performance.
	
	\item \textbf{KGMC \cite{DBLP:journals/eaai/HanKHY24}}: This technique enhances the graph by extracting keywords from user reviews using the BERT model, or the TF-IDF method or the text-rank method. If two nodes share a common keyword, an edge is established between them. For each user–item interaction, a one-hop subgraph is extracted (similar to IGMC), the target interaction edge is removed, and the modified subgraph is then processed by a relational graph convolutional network to predict the missing edge.
\end{itemize}

\subsection{Datasets}

We evaluate the algorithms on the following datasets: {Amazon\_music}\footnote{%
	\label{fn:amazon}%
	\href{https://jmcauley.ucsd.edu/data/amazon}{https://jmcauley.ucsd.edu/data/amazon}%
}, {Amazon\_office}\footref{fn:amazon}, {Amazon\_movie}\footref{fn:amazon}, and {Amazon\_tools}\footref{fn:amazon}. A five-core filtering strategy is applied, meaning users and items with fewer than five interactions are removed. The statistical analysis of the datasets used is presented in Table \ref{tab:datasets}.

	\begin{table}[H]
		\centering
		\caption{Statistical overview of the datasets}
		\label{tab:datasets}
		\renewcommand{\arraystretch}{2}
		\begin{tabular}{|c|c|c|c|c|c|}
			\hline
			Dataset & Number of users & Number of items & Number of interactions & Density  \\
			\hline
			{Amazon\_music\footref{fn:amazon}} & 27,530 & 10,620 & 231,392 & {0.079\%}  \\ 
			\hline
			{Amazon\_office\footref{fn:amazon}} & 4,905 & 2,420 & 53,258 & {0.448\%}  \\ 
			\hline
			{Amazon\_movie\footref{fn:amazon}} & 7,559 & 5,744 & 52,326 & {0.121\%} \\
			\hline
			{Amazon\_tools\footref{fn:amazon}} & 16,638 & 10,217 & 134,476 & {0.079\%} \\
			\hline
		\end{tabular}
	\end{table}
	
\subsection{Experimental settings}

The initial dataset is divided into three parts: training, validation, and test, with proportions of 60\%, 20\%, and 20\%, respectively. Hyperparameters are tuned using the validation set, and final results are obtained on the test data.  
The batch size for all models is set to 128. The learning rate for all baseline models and our proposed model is set to 0.001. The weight decay parameter  in our model is 0.09.  

In the loss function, the coefficients for the ${KL}$ term and the contrastive learning loss are set to  0.001. Additionally, for the reconstruction loss coefficient of the entire subgraph, the values 0.001, 0.005, and 0.01 are applied. 
The embedding size for each model in the proposed approach is 64, and after concatenation, the final embedding size for each node becomes 128.

\subsection{Performance comparison}

In this section, we compare the performance of the proposed method against the baseline methods according to the evaluation metrics mentioned earlier.

	\begin{table}
		\centering
		\caption{Comparison of the Proposed Method with Baseline Models for the {Amazon\_music} Dataset}
		\label{tab:topk-comparison1}
		\renewcommand{\arraystretch}{2}
			\begin{tabular}{|c|c|p{2cm}|p{2cm}|p{2cm}|p{2cm}|}
				\hline
				{Model} & {RMSE} & {NDCG@10-rating} & {MRR@10-rating} & {NDCG@10-ranking} & {MRR@10-ranking} \\
				\hline
				{IGMC \cite{DBLP:conf/iclr/ZhangC20}} & 0.881 & 0.088 & 0.139 & 0.146 & 0.174 \\ 
				\hline
				{SGL \cite{DBLP:conf/sigir/WuWF0CLX21}} & 0.985 & 0.091 & 0.111 & 0.128 & 0.158 \\ 
				\hline
				{SimGCL \cite{DBLP:conf/sigir/YuY00CN22}} & 0.971 & 0.094 & 0.122 & 0.155 & 0.170 \\ 
				\hline
				{KGMC \cite{DBLP:journals/eaai/HanKHY24}} & \underline{0.879} & \underline{0.098} & \underline{0.145} & \underline{0.222} & \underline{0.198} \\
				\hline
				{MCCL} & \textbf{0.875} & \textbf{0.112} & \textbf{0.166} & \textbf{0.284} & \textbf{0.254} \\
				\hline
				{Improvement} & 0.4 \% & 14 \% & 14 \% & 27 \% & 28 \% \\
				\hline
			\end{tabular}
	\end{table}

	\begin{table}
		\centering
		\caption{Comparison of the Proposed Method with Baseline Models for the {Amazon\_office} Dataset}
		\label{tab:topk-comparison2}
		\renewcommand{\arraystretch}{2}
			\begin{tabular}{|c|c|p{2cm}|p{2cm}|p{2cm}|p{2cm}|}
				\hline
				{Model} & {RMSE} & {NDCG@10-rating} & {MRR@10-rating} & {NDCG@10-ranking} & {MRR@10-ranking} \\
				\hline
				{IGMC \cite{DBLP:conf/iclr/ZhangC20}} & \underline{0.852} & 0.016 & 0.019 & 0.167 & 0.134 \\ 
				\hline
				{SGL \cite{DBLP:conf/sigir/WuWF0CLX21}} & 0.950 & 0.016 & 0.020 & 0.187 & 0.133 \\ 
				\hline
				{SimGCL \cite{DBLP:conf/sigir/YuY00CN22}} & 0.931 & \textbf{\underline{0.038}} & \underline{0.042} & \textbf{\underline{0.249}} & \textbf{\underline{0.222}} \\
				\hline
				{KGMC \cite{DBLP:journals/eaai/HanKHY24}} & 0.860 & 0.018 & 0.019 & 0.219 & 0.171 \\
				\hline
				{MCCL} & \textbf{0.845} & 0.036 & 0.040 & 0.234 & 0.210 \\
				\hline
				{Improvement} & 0.7 \% & -5 \% & -4 \% & -6 \% & -6 \% \\
				\hline
			\end{tabular}
	\end{table}

	\begin{table}
		\centering
		\caption{Comparison of the Proposed Method with Baseline Models for the {Amazon\_movie} Dataset}
		\label{tab:topk-comparison3}
		\renewcommand{\arraystretch}{2}
			\begin{tabular}{|c|c|p{2cm}|p{2cm}|p{2cm}|p{2cm}|}
				\hline
				{Model} & {RMSE} & {NDCG@10-rating} & {MRR@10-rating} & {NDCG@10-ranking} & {MRR@10-ranking} \\
				\hline
				{IGMC \cite{DBLP:conf/iclr/ZhangC20}} & 0.859 & 0.082 & 0.096 & 0.169 & 0.156 \\ 
				\hline
				{SGL \cite{DBLP:conf/sigir/WuWF0CLX21}} & 0.972 & 0.093 & 0.112 & 0.191 & 0.170 \\
				\hline
				{SimGCL \cite{DBLP:conf/sigir/YuY00CN22}} & 0.964 & \underline{0.106} & \underline{0.131} & \underline{0.295} & \underline{0.271} \\
				\hline
				{KGMC \cite{DBLP:journals/eaai/HanKHY24}} & \underline{0.856} & 0.104 & 0.125 & 0.223 & 0.198 \\
				\hline
				{MCCL} & \textbf{0.850} & \textbf{0.123} & \textbf{0.151}  & \textbf{0.397} & \textbf{0.370} \\
				\hline
				{Improvement} & 0.6 \% & 16 \% & 15 \% & 34 \%  & 36 \% \\
				\hline
			\end{tabular}
	\end{table}

	\begin{table}
		\centering
		\caption{Comparison of the Proposed Method with Baseline Models for the {Amazon\_tools} Dataset}
		\label{tab:topk-comparison4}
		\renewcommand{\arraystretch}{2}
			\begin{tabular}{|c|c|p{2cm}|p{2cm}|p{2cm}|p{2cm}|}
				\hline
				{Model} & {RMSE} & {NDCG@10-rating} & {MRR@10-rating} & {NDCG@10-ranking} & {MRR@10-ranking} \\
				\hline
				{IGMC \cite{DBLP:conf/iclr/ZhangC20}} & \underline{0.967} & 0.016 & 0.024 & 0.150 & 0.120 \\ 
				\hline
				{SGL \cite{DBLP:conf/sigir/WuWF0CLX21}} & 1.044 & 0.020 & 0.024 & 0.162 & 0.131 \\
				\hline
				{SimGCL \cite{DBLP:conf/sigir/YuY00CN22}} & 1.021 & \underline{0.020} & \underline{0.026} & \underline{0.170} & \underline{0.148} \\
				\hline
				{KGMC \cite{DBLP:journals/eaai/HanKHY24}} & 0.970 & 0.008 & 0.019 & 0.126 & 0.105 \\
				\hline
				{MCCL} & \textbf{0.963} & \textbf{0.022} & \textbf{0.028}  & \textbf{0.192} & \textbf{0.166} \\
				\hline
				{Improvement} & 0.4 \% & 10 \% & 7 \% & 12 \% & 12 \% \\
				\hline
			\end{tabular}
	\end{table}

\textbf{Observations:}  

\begin{itemize}
\item
As shown in Tables 
\ref{tab:topk-comparison1}, \ref{tab:topk-comparison2},
\ref{tab:topk-comparison3} and \ref{tab:topk-comparison4}, MCCL improves both the 
{\em quality of predicted values} and the {\em ranking metrics}. The improvement in RMSE over IGMC and KGMC is relatively small, possibly because both baseline models compute only MSE in their loss functions, measuring the difference between predictions and actual ratings. However, our proposed method considers predictions across different views individually as well as their final representations while incorporating contrastive learning loss and $KL$ regularization.  
\item
RMSE measures {\em error in numerical predictions}, meaning small numerical changes may not significantly affect this metric. In contrast, ranking-based metrics assess {\em how well top-ranked items are correctly identified}.
Even with minor improvements in numeric accuracy, substantial changes in ranking quality result in significant improvements in ranking metrics. Our predictions better differentiate {\em preferred and non-preferred items} but do not dramatically alter numeric predictions.
\item
Baseline models leveraging contrastive learning show weaker RMSE performance across all datasets because their focus is on {\em ranking improvement rather than absolute rating prediction}. These models assign a binary 1 or 0 to interactions rather than considering user-assigned ratings.
\item
Interestingly, SimGCL outperforms our method in Amazon\_offic for ranking metrics, despite weaker RMSE performance. This dataset has {\em higher density}, which suggests that our noise-removal approach and SimGCL's noise-handling method work best in {\em denser datasets}, where more edges provide sufficient structural learning.
In higher-density datasets, the number of edges is greater, providing the model with more data for learning. In the attention mechanism we applied, this increased data availability allows the model to better identify and eliminate noises, leading to improved performance.
\end{itemize}

\section{Ablation study}

\begin{table}[H]
	\centering
	\caption{Comparison of the our model with each of its components.}
	\label{tab:combined}
	\renewcommand{\arraystretch}{1.2}
%	\scalebox{0.75}{
		\begin{tabular}{|c|c|c|c|c|c|c|}
			\hline
			 {Dataset} &  {Model} & {RMSE} & {NDCG@10-rating} & {MRR@10-rating} & {NDCG@10-ranking} & {MRR@10-ranking}\\
			\hline
			\multirow{3}{*}{{Office}} 
			& {vgae}    & $0.858$ & $0.020$ & $0.022$ & $0.155$ & $0.130$ \\ \cline{2-7}
			& {denoising}     & $0.848$ & $0.032$ & $0.036$ & $0.208$ & $0.272$ \\ \cline{2-7}
			& {MCCL}    &  $0.845$ & $0.036$ & $0.040$ & $0.234$ & $0.210$ \\ \cline{2-7}
			\hline
			\multirow{3}{*}{{Movie}}         
			& {vgae}   & $0.882$ & $0.050$ & $0.066$ & $0.144$ & $0.112$ \\ \cline{2-7}
			& {denoising}     & $0.850$ & $0.114$ & $0.134$ & $0.331$ & $0.339$ \\ \cline{2-7}
			& {MCCL}    & $0.851$ &  $0.123$ & $0.151$ & $0.397$ &  $0.370$ \\ \cline{2-7}
			\hline
			\multirow{3}{*}{{music}}         
			& {vgae}   & $0.880$ & $0.053$ & $0.077$ & $0.128$ & $0.110$ \\ \cline{2-7}
			& {denoising}     & $0.875$ & $0.129$ & $0.175$ & $0.301$ & $0.279$ \\ \cline{2-7}
			& {MCCL} & \textbf{$0.875$} & \textbf{$0.112$} & \textbf{$0.166$} & \textbf{$0.284$} & \textbf{$0.254$} \\ \cline{2-7}
			\hline
			\multirow{3}{*}{{tools}}         
			& {vgae}   & $0.972$ & $0.009$ & $0.016$ & $0.109$ & $0.083$ \\ \cline{2-7}
			& {denoising}     & $0.963$ & $0.020$ & $0.027$ & $0.185$ & $0.158$ \\ \cline{2-7}
			& {MCCL} & $0.963$ & \textbf{$0.022$} & \textbf{$0.028$}  & \textbf{$0.192$} & \textbf{$0.166$} \\
			\hline
		\end{tabular}
%	}
\end{table}

Table \ref{tab:combined} compares the results of 
each component of our model used in contrastive learning
%our model for  of its individual components, excluding contrastive learning.
%We obtained results for each model component used in contrastive learning on the mentioned datasets. 
In this table, {\em vgae} represents Variational Graph Autoencoder, {\em denoising} refers to graph denoising mechanism using attention, and {\em MCCL} is the complete model incorporating contrastive learning.
As observed, {\em denoising} outperforms {\em vgae} across all datasets. Additionally, except for {Amazon\_music}, the complete model outperforms each of its individual components.
When comparing the individual components of our model with existing approaches,  {\em denoising} performs as well as {MCCL} across all datasets except {Amazon\_office}, showing better results than existing methods. However, 
{\em vgae} underperforms compared to previous methods in most cases.

These results indicate that our used attention mechanism can effectively remove noise and extract crucial features from data. Our complete model demonstrates that, across all datasets except Amazon\_music, integrating features extracted from both components enhances performance.  
In Amazon\_music, the {\em denoising} model outperforms the complete model.
This could be because the {\em vgae} introduces errors, and applying contrastive learning loss negatively impacts the {\em denoising} model in this dataset. Consequently, the complete model in Amazon\_music dataset exhibits inferior results compared to the standalone  {\em denoising} model.

\subsection{Hyperparameter study}

In this section, we study the effect of three main hyperparameters of our model, namely $\alpha$, $\beta$ and $\lambda$ coefficients, on its performance.

\subsubsection{Effect of the $\alpha$ coefficient}

\begin{figure}[H]
	\centering
	\includegraphics[width=1\textwidth, height=5cm]{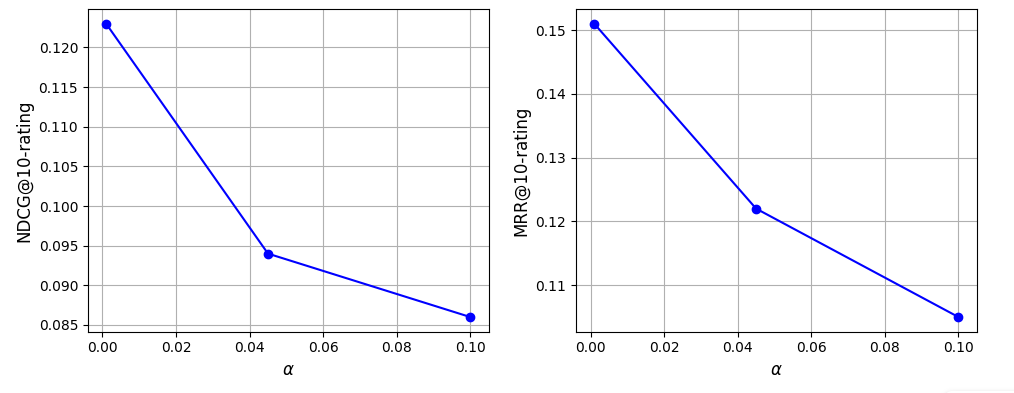}
	\caption{Impact of the $\alpha$ Parameter on the {Amazon\_movie} Dataset.}
	\label{fig:movie_alpha}
\end{figure}

\begin{figure}[H]
	\centering
	\includegraphics[width=1\textwidth, height=5cm]{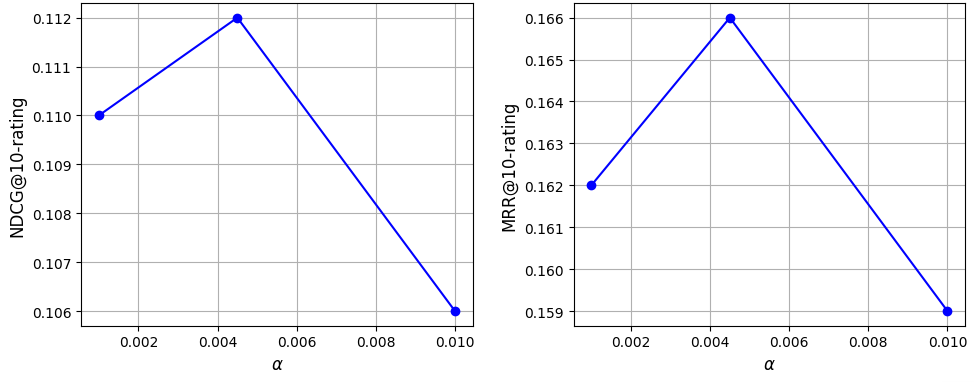}
	\caption{Impact of the $\alpha$ Parameter on the {Amazon\_music} Dataset.}
	\label{fig:music_alpha}
\end{figure}

To evaluate the impact of the full subgraph construction coefficient in the loss function for the variational graph autoencoder--denoted as $\alpha$--we test its different values: $[0.01, 0.0045, 0.001]$, on the {Amazon\_movie} and  {Amazon\_office} datasets.
For the {Amazon\_movie} dataset, the results are presented in 
Figure~\ref{fig:movie_alpha}.
For the {Amazon\_office} dataset, the results are depicted in
Figure~\ref{fig:music_alpha}.
As shown in the figures, different values of $\alpha$ yield distinct results. Among the tested values, the optimal $\alpha$ for the {Amazon\_movie} dataset is 0.001, while for the {Amazon\_music} dataset, it is 0.0045.

\subsubsection{Effect of the $\beta$ coefficient}

To evaluate the impact of the $\beta$ coefficient on the {Amazon\_movie} dataset, we test its different values: $[0.01, 0.0045, 0.001, 0.0001]$. The results are presented in Figure~\ref{fig:movie_lambda}.
As shown in the figure , different values of  $\beta$ lead to varying results. Among the tested values, the optimal $\beta$ for the {Amazon\_movie} dataset is 0.001. This coefficient balances the trade-off between divergence from a standard distribution and reliance on input data.

Increasing $\beta$ makes the model more inclined toward a standard Gaussian distribution, which may reduce dependency on input data. Conversely, lower values allow the model to focus more on input data, increasing the risk of overfitting. In this scenario, setting $\beta$ lower than 0.001 causes the model to place excessive focus on input data, leading to overfitting issues.

\begin{figure}[H]
	\centering
	\includegraphics[width=1\textwidth, height=5cm]{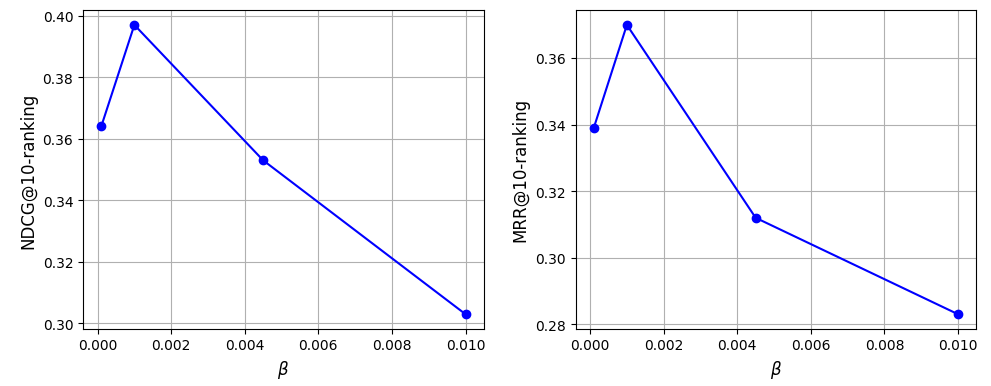}
	\caption{Impact of the $\beta$ Parameter on the {Amazon\_movie} Dataset.}
	\label{fig:movie_lambda}
\end{figure}

\subsubsection{Effect of the $\lambda$ coefficient}

To analyze the impact of the $\lambda$ coefficient on the {Amazon\_movie} dataset, we test its different values: $[0.003, 0.002, 0.001, 0.0001]$. The results are presented in Figure \ref{fig:movie_lambda}.
As illustrated in the figure, varying $\lambda$ produces different results. Among the tested values, the optimal $\lambda$ for the {Amazon\_movie} dataset is 0.001.
Initially, decreasing this coefficient reduces the impact of contrastive loss, allowing the model to focus more on its main objective while preventing excessive regularization, which could degrade performance. However, further reduction in $\lambda$ weakens the results, as the views fail to effectively reinforce each other.

\begin{figure}[H]
	\centering
	\includegraphics[width=1\textwidth, height=5cm]{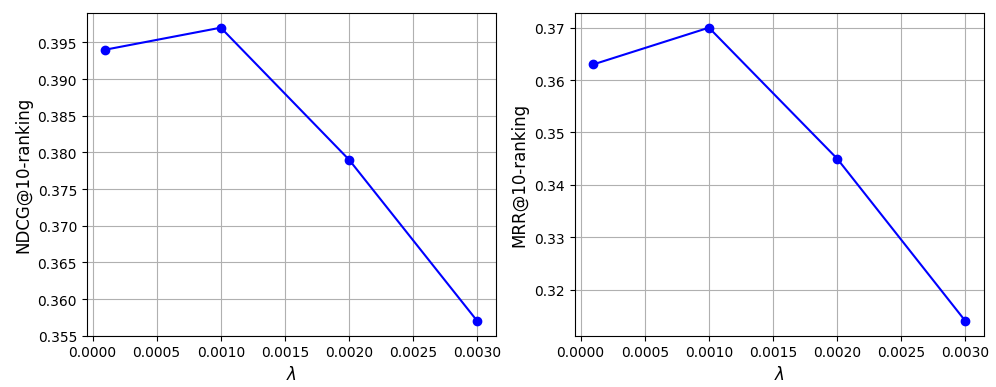}
	\caption{Impact of the $\lambda$ Parameter on the {Amazon\_movie} Dataset.}
	\label{fig:movie_lambda}
\end{figure}

\section{Conclusion}

In this paper, we addressed the problem of matrix completion in recommender systems. Our proposed method first extracted local subgraphs for each interaction. We then employed a deep learning framework that constructed two distinct subgraph views: one generated using a Variational Graph Autoencoder (VGAE) and the other utilizing a denoising model. The denoising model leveraged a Relational Graph Convolutional Network (RGCN), where attention weights were computed via a Multi-Layer Perceptron (MLP), normalized, and subsequently used for message passing in the next layer. We evaluated our method on multiple real-world datasets using both ranking quality and rating prediction accuracy metrics. The experimental results demonstrated that our approach outperformed state-of-the-art models.

\section*{Acknowledgment}

This work is supported by the Iran National Science Foundation (INSF)
under project No.4034377.

\bibliographystyle{plain}
\bibliography{references} % see references.bib for bibliography management

\end{document}